\documentclass[twocolumn]{aa}
\usepackage{graphicx}
\begin{document}
   \title{Photometric observations from theoretical flip-flop models} 

   \author{H. Korhonen \and D. Elstner}

   \offprints{H. Korhonen}

   \institute{Astrophysikalisches Institut Potsdam, An der Sternwarte 16,
              D-14882 Potsdam, Germany \\
              \email{hkorhonen@aip.de, delstner@aip.de}}

   \date{Received ; accepted}

\abstract{
Some active stars show a so-called flip-flop phenomenon in which the main spot 
activity periodically switches between two active longitudes that are
180\degr\ apart. In this paper we study the flip-flop phenomenon by converting
results from dynamo calculations into long-term synthetic photometric
observations, which are then compared to the real stellar observations. We
show that similar activity patterns as obtained from flip-flop dynamo
calculations, can also be seen in the observations. The long-term light-curve
behaviour seen in the synthesised data can be used for finding new stars
exhibiting the flip-flop phenomenon.  
}
\authorrunning{Korhonen and Elstner}
\titlerunning{Observations from the theoretical flip-flop models}
   \maketitle
%

\section{Introduction}

In many active stars the spots concentrate on two permanent active
longitudes which are $180^{\circ}$ apart. In some of these stars the
dominant part of the spot activity changes the longitude every few years.
This so-called flip-flop phenomenon was first reported by Jetsu et al.\ 
(\cite{jetsu1}, \cite{jetsu2}) in the single, late type giant FK~Com. 
Berdyugina \& Tuominen (\cite{ber_tuo}) reported periodic flip-flops between 
permanent active longitudes in four RS~CVn binaries. Their results were
confirmed in the case of II~Peg by Rodon{\`o} et al.~(\cite{rod}). The
persistent active longitude structures and flipping between two active
longitudes have also been reported over the years based on photometric
observation (e.g.\ Berdyugina et al. \cite{ber_lqhya}; Korhonen et
al. \cite{kor4}; J{\"a}rvinen et al \cite{jarv}) and on Doppler images
(Berdyugina et al. \cite{ber2}; Korhonen et al. \cite{kor_ff}). After its
discovery in cool stars, the flip-flop phenomenon has also been reported in
the Sun (Berdyugina \& Usoskin \cite{ber_sun}). A review on the flip-flop
phenomenon in cool stars and the Sun is given by Berdyugina~(\cite{ber_ff}).

In order to explain this phenomenon, a non-axisymmetric dynamo component, 
giving rise to two permanent active longitudes 180\degr\ apart, is needed
together with an oscillating axisymmetric magnetic field. Fluri \& Berdyugina
(\cite{flu_ber}) suggest also another possibility with a combination of
stationary axisymmetric and varying non-axisymmetric
components. Unfortunately, no simple dynamo mechanism is yet known for such a
configuration. There are models with anisotropic $\alpha$-effect or a weak
differential rotation, which could produce non-axisymmetric components, but
only recently Moss (\cite{moss1}, \cite{moss2}) reported coexisting mixed
components with a differential rotation that depends on the distance to the
rotation axis. This differential rotation configuration is a very plausible
state for fast rotators. Also, a weak non-axisymmetric field coexisting with a
dominant axisymmetric field, assuming a solar-like rotation law, was found by
Moss (\cite{moss99}). This solution was not analysed for flip-flops, as the
possibility of them being present on the Sun was not being discussed at that
time. Flip-flop solutions for a rotation law similar to the solar one
and anisotropic $\alpha$ have also been found by Elstner \& Korhonen
(\cite{elstner}).

According to the calculations by Elstner \& Korhonen (\cite{elstner}) the 
shift of the spots in a flip-flop event is 180\degr\ only in some cases, 
mainly for the stars with thin convective zones. In stars with thick 
convective zones they found a shift that is close to 90\degr. Similar
results were reported by Moss (\cite{moss2}). Recently, Ol{\'a}h et
al. (\cite{olah}) re-analysed some of the old photometric observations of FK
Com and found a flip-flop event in which spots on both active longitudes
vanished briefly, and one of the new spots appeared on an old active longitude
and the other one 90\degr\ away from that. This is the first evidence
suggesting that flip-flops where the spots shift only by 90\degr can also
occur.

In this paper we use the model calculations presented by Elstner \& Korhonen 
(\cite{elstner}) and convert them into synthetic photometric
observations. This is used to investigate the expected long-term photometric
behaviour of active stars showing the flip-flop phenomenon. Hopefully, this
will help us in identifying new stars exhibiting this intriguing
phenomenon. At the moment only few stars showing it are known and no
statistically significant correlation between the stellar parameters and the
flip-flop phenomenon can therefore be carried out.

\section{Model}

The model consists of a turbulent fluid in a spherical shell of inner radius
$r_{\rm in}$ and outer radius $r_{\rm out}$.

We solve the induction equation 
\begin{equation}
{\partial \langle\vec{B}\rangle \over \partial t} = 
{\rm curl}(\alpha_q\circ \langle\vec{B}\rangle
-{\eta}_{\rm T} {\rm {curl}}\langle{\vec{B}}\rangle), 
\label{eq:1}
\end{equation} 
in spherical coordinates (${\rm r,\theta,\varphi}$) for an
$\alpha^2\Omega$-dynamo. Here we used a solar-like rotation law in the
corotating frame of the core
\begin{eqnarray}
\lefteqn{\Omega(r,\theta)= 
{1\over 2}\Omega_0\left[1+{\rm erf}\left({r-r_{c} \over d_1}\right)\right]
(\Omega_{s} -\Omega_{c})}
\label{eqomeg}
\end{eqnarray} 
where $ \Omega_s=\Omega_{eq}- a {\rm cos}^2\theta $ with 
$\Omega_{eq}/2\pi=460.7{\rm nHz}$, $\Omega_c/2\pi=432.8{\rm nHz}$ and  
$a/2\pi=125.82{\rm nHz}$ are used. Normalising length with stellar radius 
$R_\star$ and time with diffusion time $t_d=R_\star^2/\eta$ we can define the
dimensionless dynamo numbers 
\begin{equation}
C_\alpha=\alpha_0 \cdot R_\star/\eta 
\label{Calpha}
\end{equation} 
and
\begin{equation} 
C_\Omega=(\Omega_{eq}-\Omega_{c}) \cdot R_\star^2/\eta . 
\end{equation} 
For ${ R}_\star={ R}_\odot$, the diffusivity $\eta=5 \cdot 10^{12} {\rm cm^2
  s^{-1}}$ and $\Omega_{0}=1$ we get $C_\Omega$=172. Notice that a definition
with $  \Omega_{eq}-\Omega_{pol}$ gives  $C_{\Omega}=780$. Because of these
different definitions of $C_{\Omega}$ in the literature we describe the
strength of the differential rotation  in our models with the value of
$\Omega_{0}$.
 
In order to identify the lifetime and the maximal possible pole to equator
difference of the angular velocity for a flip-flop solution also for models
with isotropic $\alpha$, we performed several calculations with models similar
to those presented in Moss (\cite{moss2}). Here we used the differential
rotation law (Eq. \ref{eqomeg}) with $ \Omega_s=  - a r^2{\rm sin}^2\theta$
and $ \Omega_c=0$. The same normalisation was used.

Only the symmetric part  
\begin{eqnarray}
\lefteqn{\alpha_{rr}=\alpha_0 {\rm cos}\theta(1.-2{\rm cos}^2\theta)}
\nonumber \\ 
\lefteqn{\alpha_{\theta\theta}=
\alpha_0 {\rm cos}\theta(1.-2{\rm sin}^2\theta)} \nonumber \\ 
\lefteqn{\alpha_{\varphi\varphi}=\alpha_0 {\rm cos}\theta} \nonumber \\
\lefteqn{\alpha_{r\theta}=\alpha_{\theta r}= 2 \alpha_0 {\rm
    cos}^2\theta{\rm sin}\theta}  
\end{eqnarray} 
of the $\alpha$-tensor is included (cf. R{\"u}diger et al \cite{rued}). For
the isotropic model $\alpha_{r\theta}=\alpha_{\theta r}=0$ and the diagonal
terms $\alpha_{ii}=\alpha_0 {\rm cos}\theta $. In order to saturate the dynamo
we choose a local quenching of
\begin{equation}
\alpha_q={\alpha \over 1+\vec{B}^2/\vec{B}_{\rm eq}^2}$$.
\label{eq:quench}
\end{equation}

The inner boundary is a perfect conductor at $r_{\rm in}=0.3$ and the outer
boundary resembles a vacuum condition, by including an outer region up to 1.2
stellar radii into the computational grid with 10 times higher diffusivity. At
the very outer part the pseudo vacuum condition (tangential component of the
magnetic field and vertical component of the electric field vanish on the
surface)  is used. In order to see the influence of the thickness of the
convection zone we have chosen $r_{\rm c}=0.4$ for a thick (results shown in
Fig.~\ref{model}a) and $r_{\rm c}=0.7$ for a thin (Fig.~\ref{model}b)
convection zone. Below the convection zone from $r_{\rm in}$ to $r_{\rm c}$
the diffusivity was reduced by a factor 1000 and $\alpha$ was set to zero. The
critical $C_\alpha$ and periods for the models are given in
Table~\ref{linear}.

In all the models we used a normalised $\eta=0.5 $, $C_\alpha$ and $\Omega_0$ 
for the thin and thick convection zone models are given in Table \ref{param}
and the parameters for the model used in Fig.~\ref{enermota} are
$\Omega_0=-0.3$, $r_{\rm in}=0.2$ and  $r_{\rm c}=0.3$ $C_\alpha=8$. 

\begin{table}
\caption{Critical $C_\alpha$ for the symmetric and antisymmetric
    components with azimuthal wavenumbers m=0,1. The second number gives the
    oscillation period (for m=0) or the migration period (for m=1) in
    diffusion times (30 years).} 
\begin{tabular}{lllccccc}\hline
name  &      S0       &     S1     &    A0       &    A1      \\
\hline
thick &  9.7  0.24    &  9.0 0.15  &  9.7  0.24  &  9.0  0.15  \\
thin  & 11.9  0.18    & 10.2 0.27  & 11.9  0.18  & 10.2  0.27  \\
\hline
\end{tabular}
\label{linear}
\end{table}

\section{From the model to observations}

For converting the possible spot pattern from the model calculation into
synthetic photometric observations, we first have to decide which value of the
magnetic pressure on the stellar surface results in a spot. For doing this a
three temperature model was chosen in which the values of magnetic pressure
that are $\ge 70$~\% of the maximum value are considered to form the ``umbra'' 
and the values $< 70$~\% and $\ge 30$~\% of the maximum form the
``penumbra''. The values $< 30$~\% of the maximum denote the unspotted
surface. For investigating the long-term changes in the spot strength the
maximum value of the magnetic pressure was taken from the whole run, not from
the individual maps.

\begin{figure}
   \centering
   \includegraphics{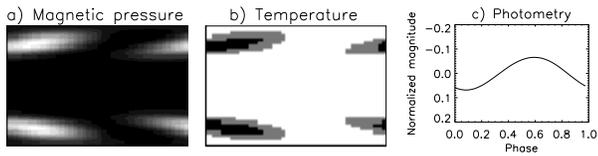}
   \caption{From the model calculations to the photometric observations. a)  
The magnetic pressure on the surface obtained from the model calculations.
The brightest areas correspond to the largest magnetic pressure. b) The 
temperature map showing the spot configuration corresponding to the magnetic 
pressure distribution. In the map the black area has a temperature of 3500~K,
grey corresponds to the temperature of 4250~K and white is the unspotted 
surface with the temperature of 5000~K. c) The normalised synthetic
light-curve corresponding to the temperature map.}
   \label{intro}
   \end{figure}

A typical star showing flip-flops is a cool giant or a zero age main sequence
object. For describing the realistic spot temperatures on such stars, 5000~K
was chosen as the unspotted surface temperature and 3500~K \& 4250~K as the
umbral and penumbral temperatures, respectively. After the assignment of the
spot temperatures to the magnetic pressure maps, synthetic light-curves were
calculated from the maps. The limb-darkening coefficient from Al-Naimyi
(\cite{aln}) for 5000~K at the central wavelength of the Johnson V band was
used for all the three temperatures. Fig.~\ref{intro} shows the magnetic
pressure map obtained from the dynamo calculations, the corresponding spot
configuration and the synthetic light-curve calculated from the temperature
map.

\section{Results}

The results from the thick and thin convection zone flip-flop models, that
were first discussed in Elstner \& Korhonen (\cite{elstner}), have been
converted into light-curves as described in the previous section. The model
parameters are given in Table~\ref{param}.

\begin{table}
\caption{The parameters for the models investigated in this paper. Table
  gives the name used for the model in the text, location of the inner
  boundary, the strength of the differential rotation (for solar differential
  rotation $\Omega_0=1$), dynamo-number $C_{\alpha}$ (Eq.~\ref{Calpha}), the
  energy density in equipartition units E0 for the component $m=0$ and E1 for
  $m=1$, the period of the oscillation in diffusion times and the migration
  period, also in diffusion times.}
\begin{tabular}{lllccccc}\hline
name & $r_{\rm in}$ & $\Omega_0$ & $C_{\alpha}$ &   E0 & E1 & P0 & P1 \\ 
\hline
thick & 0.4    &      0.12  &    10      &   0.06   &  0.06    &  0.3   & 0.2
\\ 
thin & 0.7    &      0.11  &    21      &   0.1    &  0.4     &  0.23  & 0.46
\\ 
\hline
\end{tabular}
\label{param}
\end{table}

\begin{figure}
   \centering
   \includegraphics{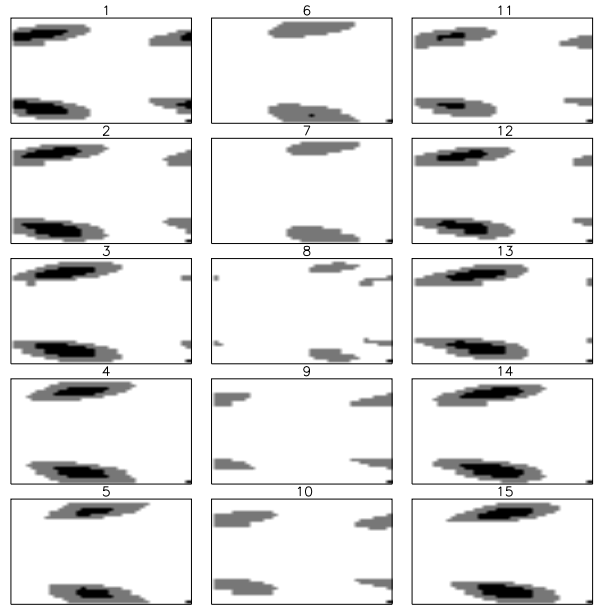}
   \caption{Temperature maps in a cylindrical projection showing a migrating
   spot pattern and a flip-flop event for the thick model. The colour coding
   is the same as in Fig.~\ref{intro}b. Each maps has 500 time steps in
   between them. Note that the small spot in the lower right corner is for 
   enabling the correct plotting of the colours.}
   \label{maps}
   \end{figure}

In Fig.~\ref{maps} an example of a sequence of temperature maps exhibiting a
migrating spot pattern and a flip-flop event is shown. The features are
symmetric with respect to the equator. It is clearly seen that in this thick
convection zone model the spot shift in the flip-flop is $\sim 90^{\circ}$ and
not 180\degr, as is more commonly seen in the observations. 

\begin{figure*}
  \setlength{\unitlength}{1mm}          
  \begin{picture}(0,70) 
    \put(10,65){\bf a)}
    \put(20,33){\begin{picture}(0,0) \includegraphics{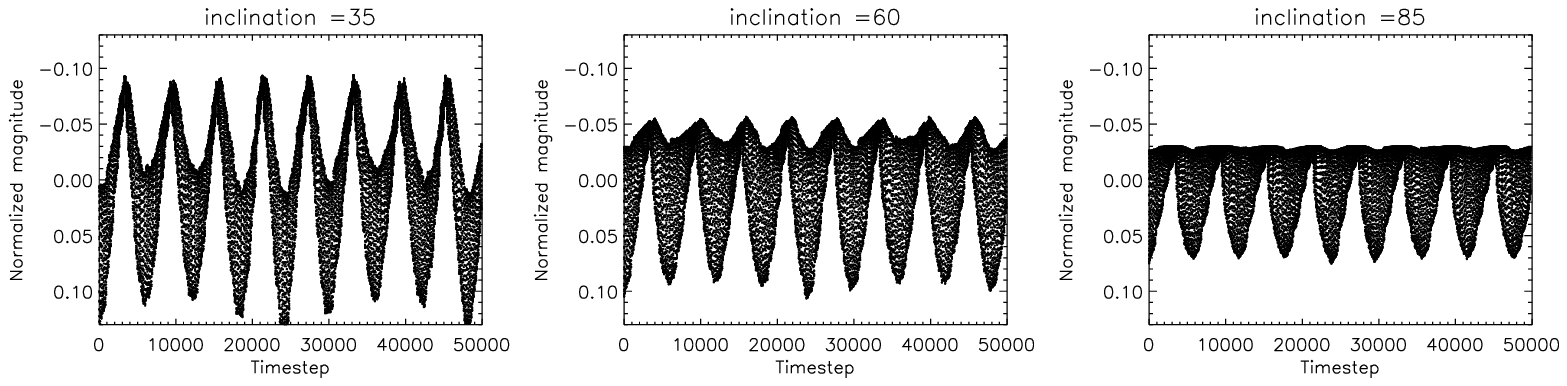} \end{picture}}
    \put(10,30){\bf b)}
    \put(20,-2){\begin{picture}(0,0) \includegraphics{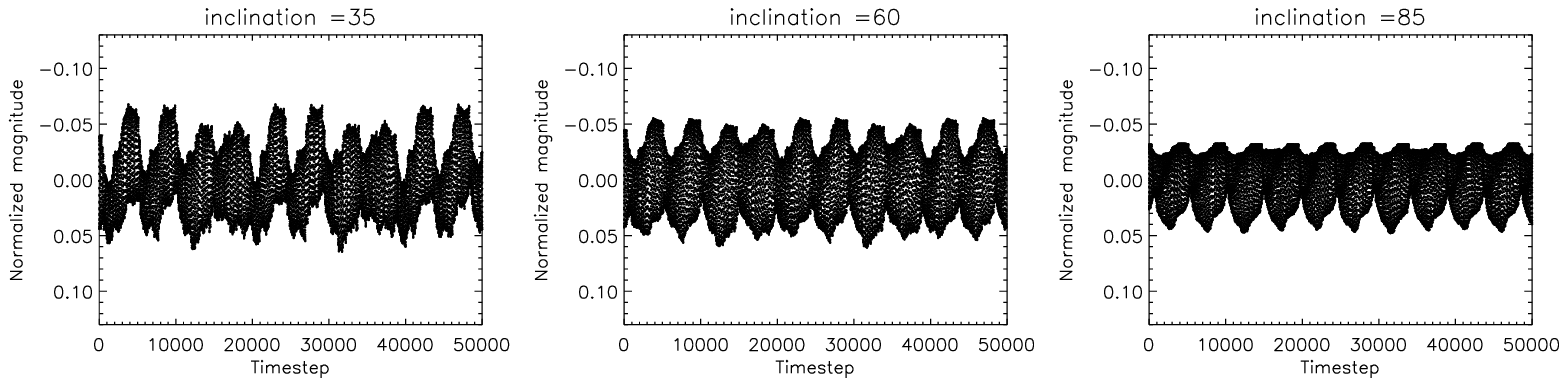} \end{picture}}
  \end{picture}
  \caption{The normalised magnitudes, calculated from the thick (a) and thin 
    (b) convection zone flip-flop models, plotted against the timestep in the
    model for three different inclination angles. The series consists of 
    synthetic light-curves calculated from 500 surface maps with 100 timesteps 
    in between each map. The length of the whole series is thus slightly longer
    than 10~000 days ($\sim$28 years). The starting time for the series is 
    around 80 convective turn-over times. The magnitudes have been normalised
    to the mean magnitude of the whole set.} 
   \label{model}
   \end{figure*}

For investigating the long-term photometric behaviour obtained from the models,
calculations were done starting around 80 diffusion times, running 50000
timesteps (one timestep is $1.9 \cdot 10^{-5}$ diffusion times or
approximately 1/5 days for our models) and calculating a map of magnetic
pressure at the surface every 100 steps. These maps were then converted into
temperature maps and synthetic light-curves, and the light-curves were plotted
against time to see the long-term behaviour. Figs.~\ref{model}a \& b show the
calculated light-curve behaviour for three different inclination angles for
the thick and thin models, respectively.

In the case of the thick model (Fig.~\ref{model}a) we see that the minimum
magnitude is very strongly modulated during the flip-flop cycle, whereas the
maximum magnitude is much less affected, except when viewed from small
inclinations. The minimum and maximum in the maximum magnitude occur around 
the time of the minimum and maximum in the minimum magnitude. This behaviour
is the same as seen by Fluri \& Berdyugina (\cite{flu_ber}) in their first case
(sign-change of the axisymmetric component). The inclination of the star
affects mainly the amplitude of the variation, as it determines how close the
spots are to the centre of the visible disk (location of the maximum effect on
the light-curve). In the case of high-latitude spots, as seen in these models,
this effect totally dominates over the other inclination effect, i.e. how much
of the spots on the ``southern'' hemisphere are visible. The changes in the
inclination affect the behaviour of the maximum magnitude more strongly than
the minimum magnitude. 

The behaviour seen in the thin convection zone model (Fig.~\ref{model}b) is
different from the one seen in the thick model. The amplitude of the variation
is much smaller because the active longitudes are further apart than in the 
thick case. It is also worth noting, that in this case the variation in the
maximum and minimum magnitudes is very different from the thick case; here the
maximum of the maximum magnitude occurs near the minimum of the minimum
magnitude. There is a small shift towards the later timesteps for the maximum
of the maximum magnitude in comparison to the minimum of the minimum magnitude.

\section{Discussion}

\subsection{Dependence of the solution on the model parameters}

\begin{figure}
\centering
   \resizebox{4.2cm}{!}{\includegraphics{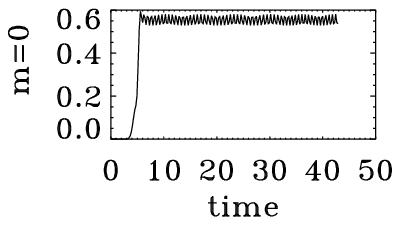}}
   \resizebox{4.2cm}{!}{\includegraphics{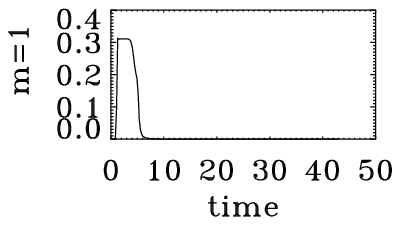}}
\caption{Magnetic energy in axisymmetric (m=0) and non-axisymmetric (m=1)
   components for a model with $r_{c}=0.3$ and isotropic $\alpha$. The pole to
   equator difference of $\Omega$ is 30\% of the solar value. The mixed 
   component flip-flop solution survives only 5 diffusion times.}
\label{enermota}
\end{figure}

The differential rotation needed for a flip-flop solution with a period of
about 5 years is rather small. For a solar sized star the latitudinal
difference in $\Omega$ ($\Delta\Omega$) should be about 10\% of the solar
value, independent of the global rotation. For $R_{\ast}=2R_{\odot}$ the
$\Delta\Omega$  can be about 40\%. The situation seems similar in all cases,
i.e. for isotropic $\alpha$ tensor used together with a differential rotation
depending on the distance to the rotation axis and with an anisotropic
$\alpha$ tensor used with both a solar-like rotation law and axis distance
dependent rotation law. An example of the time evolution of the magnetic field
energy in the m=0,1 components with isotropic $\alpha$, $\Delta\Omega$ 30\% of
the solar value, and a rotation law depending on the axis distance (cf. Moss
2005) is shown in Fig.~\ref{enermota}. As can be seen, with relatively
strong differential rotation the non-axisymmetric component, m=1, is initially
excited, but the mixed component solution survives only approximately 5
diffusion times. 

Increasing the diffusivity gives a strong flip-flop phenomenon also for
higher pole to equator differences of $\Omega$, but with a smaller flip-flop
period. The flip-flop solutions appear preferential for a positive $\alpha$ in
the northern hemisphere in the case of radially increasing $\Omega$ at the
equator. This leads to a poleward migration of the spots. Observations
indicate solar-like equatorward migration pattern in solar-like stars (see
e.g. Katsova et al.~\cite{kats}), but there is also a detection of poleward 
migration of the spots in the RS~CVn binary HR~1099 (Vogt et al.~\cite{vogt};
Strassmeier \& Bartus \cite{str_bar}). 

\subsection{Comparing the synthetic light-curves with the 
  observations} 

Many active stars show long-term light-curves where time periods with small and
large amplitude in the photometry alternate, as seen in our flip-flop
models. For the behaviour seen in the thick model (Fig.~\ref{model}a) a good
stellar counterpart is DX~Leo (see e.g. Messina \& Guinan \cite{mes}). It is
easier to find stellar counterparts for the thin case
(Fig.~\ref{model}b). Some examples, like LQ~Hya and EI~Eri, can be seen for
instance in Ol{\'a}h \& Strassmeier (\cite{olah_str}). The fact that the thin
case seems to be dominant implies that the flip-flops where the spots shift
180\degr\ are more common. 

\begin{figure*}
  \setlength{\unitlength}{1mm}          
  \begin{picture}(0,60) 
    \put(5,50){\bf a)}
    \put(10,5){\begin{picture}(0,0) \includegraphics{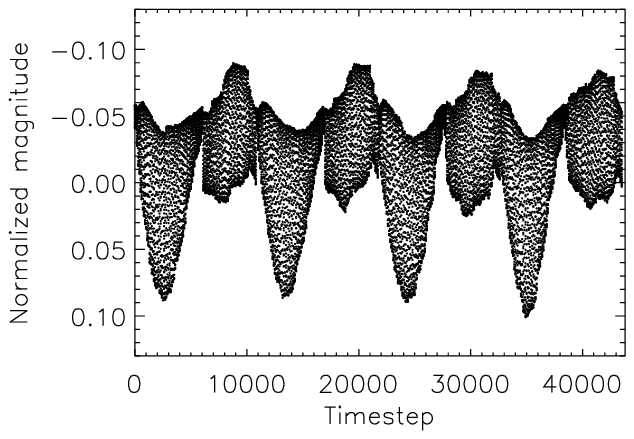} \end{picture}}
    \put(81,50){\bf b)}
    \put(85,0){\begin{picture}(0,0) \includegraphics{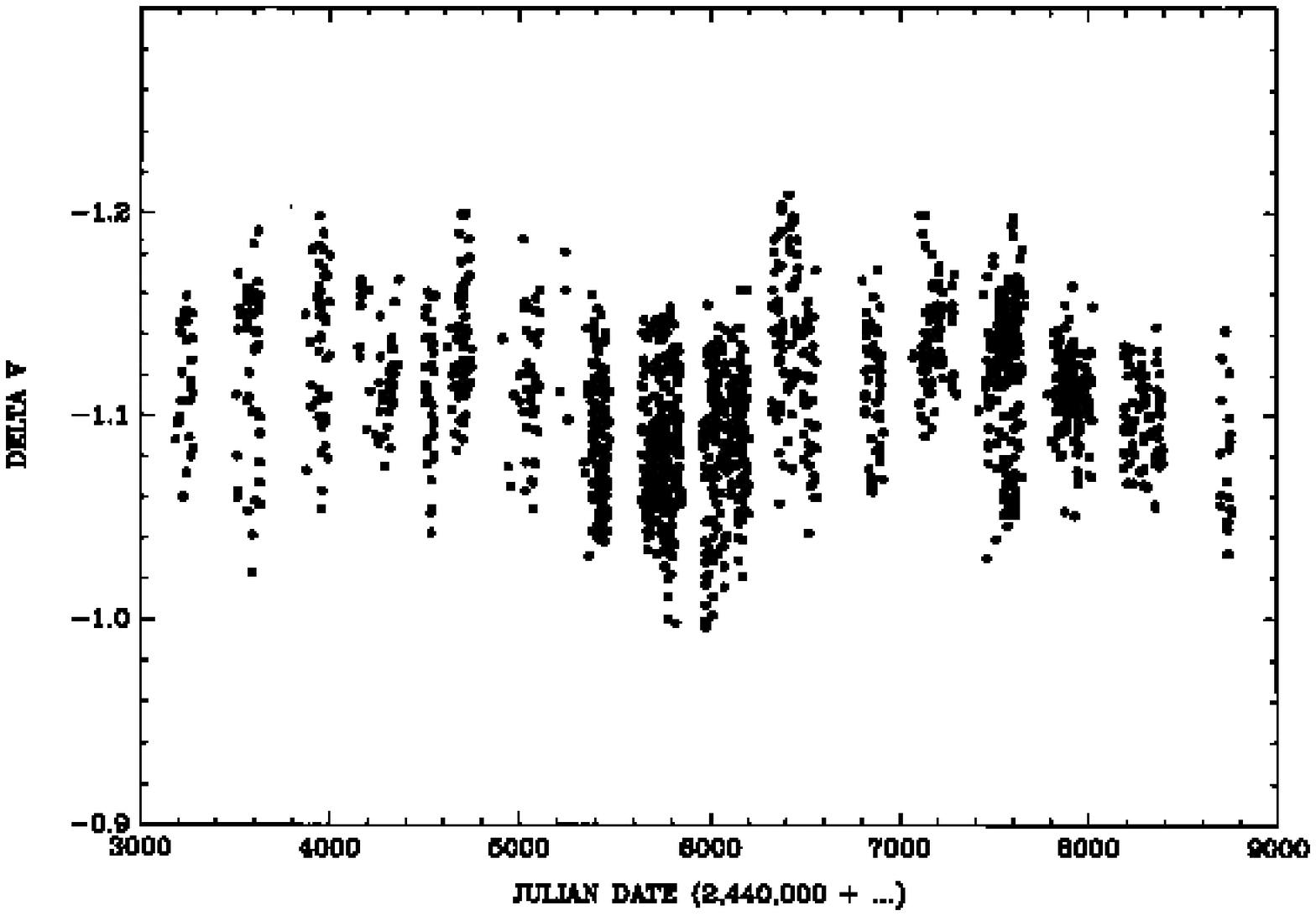} \end{picture}}
  \end{picture}
   \caption{a) The combination of the synthetic light-curves from the 90\degr\ 
    and 180\degr\ flip-flops running approximately 9000 days, every other
    flip-flop is from different model. The 180\degr\ flip-flop calculations
    were moved slightly up for a more continuous plot. b) Light-curve of 
    $\sigma$~Gem (from Henry et al.~\cite{henry}) shows indications of similar
    behaviour as seen in the panel a. Note the longer time span of the model 
    (a) in comparison to the observations (b).} 
   \label{thin_thick}
   \end{figure*}

As seen in the case of FK Com (Ol{\'a}h et al \cite{olah}), some stars can
show both 90\degr\ and 180\degr\ shifts in the spots during a flip-flop
event. For investigating what alternating 90\degr\ and 180\degr\  flip-flops
would look like in the long-term photometry, we combine the calculated
light-curves from the models showing 90\degr\ (from the thick model) and
180\degr\ (from the thin model) flip-flops, taking alternatively one 90\degr\
flip-flop and one 180\degr\ flip-flop. The result of combining the  two types
of flip-flops is shown in Fig.~\ref{thin_thick}a. The long-term light-curve
behaviour produced by this is similar, but not identical, to the second case
of Fluri \& Berdyugina (\cite{flu_ber}), which shows alternatively small and
large amplitude changes that are symmetric with respect to the mean magnitude,
i.e. the maximum of the maximum magnitude occurs at the time of the minimum of
the minimum magnitude. Our combination of the models with different spot
shifts in the flip-flop event is symmetrical only during the smaller amplitude
phase. During the larger amplitude phase the minimum of the maximum magnitude
occurs at the time of the minimum of the minimum magnitude. For example, the
light-curve of $\sigma$~Gem (see Fig.~\ref{thin_thick}b) shows indications of
this kind of behaviour. 

It is often difficult to see in the real observations the activity pattern
caused by the flip-flop behaviour. This is partly due to the years long time
series of observations needed to see the pattern and partly due to the
solar-like cyclic changes in the over-all activity level of many active stars.
Quite drastic changes in the brightness of some stars can be seen on top of
the possible flip-flop signature (see e.g. HK~Lac and IL~Hya in Ol{\'a}h \&
Strassmeier \cite{olah_str}) and these large changes are likely to mask the
patterns caused by the flip-flops.

\section{Summary}

We have synthesised photometry from Dynamo calculations exhibiting flip-flop
behaviour. This was done for investigating the long-term changes in the
photometric behaviour seen over several flip-flop cycles. On the  whole, the
activity patterns discussed in this paper imply flip-flop phenomenon and stars
showing these patterns should be further investigated for checking if they
really show flip-flops. A statistically significant sample of stars is needed
for deeper understanding of this phenomenon. Further, more effort should be
put to measuring the $\Delta\Omega$, meridional flow, and latitude migration
of the spots on different types of stars. All these parameters have important
implications for the dynamo calculations.

\begin{acknowledgements}
We would like to thank the referee Dr.\ Moss for his very useful comments on
this paper. HK acknowledges the support from the German \emph{Deut\-sche
  For\-schungs\-ge\-mein\-schaft, DFG\/} project KO~2320/1. 
\end{acknowledgements}

\end{document}